\documentclass[iop,twocolumn,apjl]{emulateapj}
\usepackage{apjfonts}

\epsscale{1.15}

\shorttitle{Kpc-Scale Binary AGNs in Stripe 82} 
\shortauthors{Fu et al.}
\journalinfo{ApJ Letters in press}
\submitted{Received 2015 October 1; accepted 2015 November 20}

\usepackage[pagebackref=false,letterpaper=true,colorlinks=true,citecolor=blue,linkcolor=blue,breaklinks=true,bookmarks=true]{hyperref}

\newcommand{\kms}{{km s$^{-1}$}}

\newcommand{\msun}{$M_{\odot}$}

\newcommand{\uJy}{$\mu$Jy}

\newcommand{\OIII}{[O\,{\sc iii}]}

\begin{document}

\title{Binary Active Galactic Nuclei in Stripe 82: Constraints on Synchronized Black Hole Accretion in Major Mergers}

\author{
Hai~Fu\altaffilmark{1}, J.~M.~Wrobel\altaffilmark{2}, A.~D.~Myers\altaffilmark{3}, S.~G.~Djorgovski\altaffilmark{4}, and Lin~Yan\altaffilmark{4}
}
\altaffiltext{1}{Department of Physics \& Astronomy, University of Iowa, Iowa City, IA 52245}
\altaffiltext{2}{National Radio Astronomy Observatory, P.O. Box O, Socorro, NM 87801}
\altaffiltext{3}{Department of Physics \& Astronomy, University of Wyoming, Laramie, WY 82071}
\altaffiltext{4}{California Institute of Technology, 1200 E. California Blvd., Pasadena, CA 91125}

\begin{abstract}
Representing simultaneous black hole accretion during a merger, binary active galactic nuclei (AGNs) could provide valuable observational constraints to models of galaxy mergers and AGN triggering. High-resolution radio interferometer imaging offers a promising method to identify a large and uniform sample of binary AGNs, because it probes a generic feature of nuclear activity and is free from dust obscuration. Our previous search yielded 52 strong candidates of kpc-scale binaries over the 92\,deg$^2$ of the Sloan Digital Sky Survey (SDSS) Stripe 82 area with 2\arcsec-resolution Very Large Array (VLA) images. Here we present 0\farcs3-resolution VLA 6~GHz observations for six candidates that have complete optical spectroscopy. The new data confirm the binary nature of four candidates and identify the other two as line-of-sight projections of radio structures from single AGNs. The four binary AGNs at $z\sim0.1$ reside in major mergers with projected separations of 4.2$-$12\,kpc. Optical spectral modeling shows that their hosts have stellar masses between $10.3<{\rm log}(M_\star/M_\odot)<11.5$ and velocity dispersions between $120<\sigma_\star<320$\,\kms. The radio emission is compact ($\lesssim0\farcs4$) and show steep spectrum ($-1.8<\alpha<-0.5$) at 6\,GHz. The host galaxy properties and the Eddington-scaled accretion rates broadly correlate with the excitation state, similar to the general radio-AGN population at low redshifts. Our estimated binary AGN fraction indicates that simultaneous accretion occurs $\geqslant$$23^{+15}_{-8}$\% of the time when a kpc-scale galaxy pair is detectable as a radio-AGN. The high duty cycle of the binary phase strongly suggests that major mergers can trigger and synchronize black hole accretion.
\end{abstract}


\section{Introduction} \label{sec:intro}

Today's massive galaxies were built upon a series of major and minor mergers. Two important observations suggest that major mergers are critical to galaxy evolution: the tight correlation between black hole mass and bulge properties in the local universe (e.g., the $M_{\bullet}$-$\sigma_{\star}$ correlation; \citealt{McConnell13}), and the striking similarity between the cosmic growth histories of galaxies and supermassive black holes (SMBHs) \citep[e.g.,][]{Zheng09}. Given that SMBHs mainly grow through accretion \citep{Yu02}, gas-rich major mergers offer an attractive theoretical framework to explain these observations. Gas-rich mergers can accomplish the transformation of galactic disks into spheroids and they can drive the development of bars that can funnel large amounts of gas to the nucleus of a galaxy in less than the dynamical timescale \citep[e.g.,][]{Barnes92}. The resulting sudden gas inflow could trigger nuclear starbursts and provides a possible fueling mechanism for SMBHs, thus initiating an episode of galaxy-SMBH co-evolution \citep[e.g.,][]{Hernquist89}. Lastly, feedback from the accreting SMBHs is needed to prevent excessive star formation by either ejecting materials from galactic disks \citep[quasar-mode; e.g.,][]{Di-Matteo05} or by maintaining the high temperatures of X-ray gas in the host halo \citep[radio-mode; e.g.,][]{Best06}. 

However, whether major mergers are the dominant mechanism that triggers black hole accretion in AGN is still under debate. While an excess of AGN is found in galaxy pairs at separations below $\sim$50~kpc \citep[e.g.,][]{Silverman11,Liu12,Koss12,Ellison13,Satyapal14}, AGN host galaxies and inactive galaxies at $z \sim 1-2$ have similar morphologies. The majority of AGN hosts at $z \sim 1-2$ are disk galaxies \citep[e.g.,][]{Cisternas11,Schawinski11a,Kocevski12,Villforth14}, but at lower redshift AGNs are more likely to be in spheroids and pairs \citep[e.g.,][]{Dunlop03,Bennert08,Koss10}. There are several possible ways to explain these apparently contradictory results: (1) major mergers can trigger AGNs, but most AGNs are triggered by minor disturbances; (2) AGNs are highly variable on timescales shorter than the lifetime of merger signatures; and (3) observations at higher redshifts are not sensitive enough to detect the low surface-brightness tidal features seen at low redshifts. 

A critical phase in a major merger is when the SMBHs in both galaxies are accreting simultaneously to produce a pair of AGNs. Such binary AGNs could provide further insights into AGN fueling mechanisms \citep[e.g.,][]{Foreman09}. Previous optical studies found an excess in quasar clustering between $10<\rm{Sep}<40\,\rm{kpc}$ \citep[e.g.,][]{Hennawi06} and an elevated AGN fraction in binary AGNs with $\rm{Sep}\lesssim10\,\rm{kpc}$ \citep[e.g.,][]{Fu12a,Liu12}. But AGNs in gas-rich mergers could be heavily obscured by dust, making them undetectable through optical spectroscopy. In addition, optically selected kpc-scale binary AGNs require confirmation with either X-ray or radio observations to prove the existence of separate central engines. For example, extended emission-line regions photoionized by a single AGN could appear as a binary AGN in fiber/longslit spectra if part of the emission-line gas is aligned with the companion galaxy \citep{Fu12a}. As a consequence, convincing cases of kpc-scale binary AGNs are rare \citep[e.g.,][]{Junkkarinen01,Komossa03,Hudson06,Koss11,Fu11b,Comerford11,Fabbiano11,Comerford15,Muller-Sanchez15} and a homogeneous sample has been lacking.

We therefore conducted a systematic search of kpc-scale binary AGNs directly from wide-field high-resolution radio imaging \citep{Fu15}. Radio-selected samples are known to have a high AGN fraction \citep[e.g.,][]{Best12}. The Very Large Array (VLA) survey of the Sloan Digital Sky Survey (SDSS) Stripe 82 field \citep{Hodge11} covers a total area of $\sim$92\,deg$^2$ with an angular resolution of 1.8\arcsec\ and a median rms noise of 52\,\uJy\,beam$^{-1}$. After removing obvious contaminants (such as projected pairs in large radio lobes), we identified 52 candidate binary AGNs. The 22 grade-A candidates are most promising because of their almost unambiguous radio-optical association, while the other 30 grade-B candidates are more likely to be projected pairs.
From a small sample that have spectra for both components, we confirmed six as kpc-scale mergers with $\Delta\,v<500$~\kms. In this {\it Letter}, we present 0\farcs3-resolution VLA observations to study the nature of the six candidates. We adopt the $\Lambda$CDM cosmology with $\Omega_{\rm m}=0.27$, $\Omega_\Lambda=0.73$, and $h=0.7$.

\section{Observations} \label{sec:obs}

\begin{figure*}[!t]
\epsscale{0.9}
\plotone{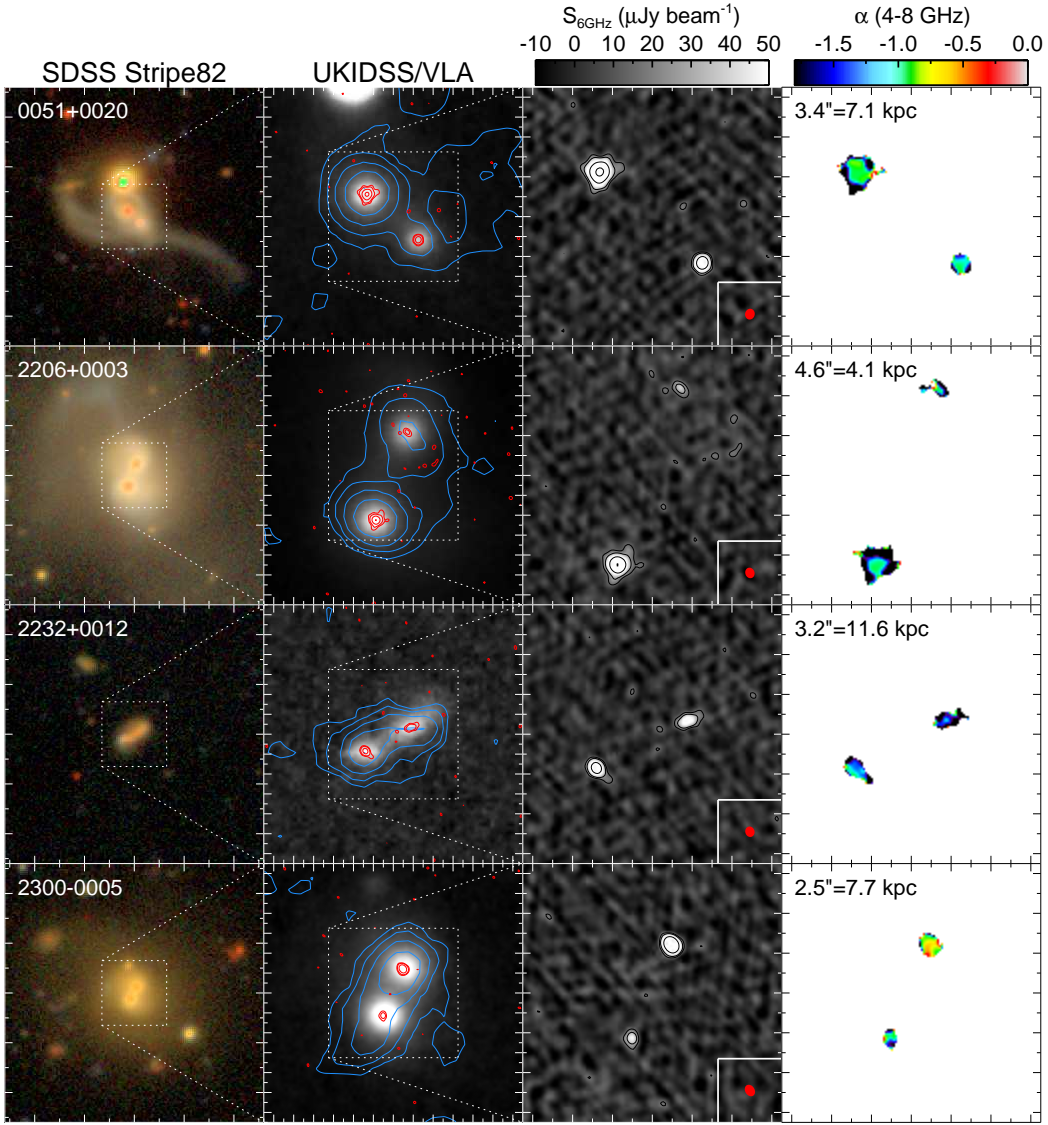}
\caption{The confirmed binary radio-AGNs. For each system, we show from left to right: a wide-field SDSS Stripe82 coadded $gri$ color image, a narrow-field UKIDSS $J$-band image overlaid with our VLA 6\,GHz A-configuration continuum map ({\it red} contours) and the 1.4\,GHz continuum map from the VLA-Stripe82 survey \citep[{\it blue} contours;][]{Hodge11}, a further zoomed-in version of the 6\,GHz continuum map and its restoring beam (the red ellipse at the lower right corner), and a 4$-$8~GHz spectral index map. The red contours in the second column and the black contours in the third column are at ($+$3, $+$6, $+$24, $+$96)$\times$$\sigma$. The lowest blue contours in the second column are at 2$\sigma$ and the levels increase exponentially to the peak value of each map. Major tickmarks are spaced in 10\arcsec\ intervals in the leftmost column and 1\arcsec\ intervals for the other columns. The projected separation of each pair is labeled in the rightmost column. N is up and E is left for all panels.
\label{fig:binaries}}
\end{figure*}

\begin{figure}[!t]
\plotone{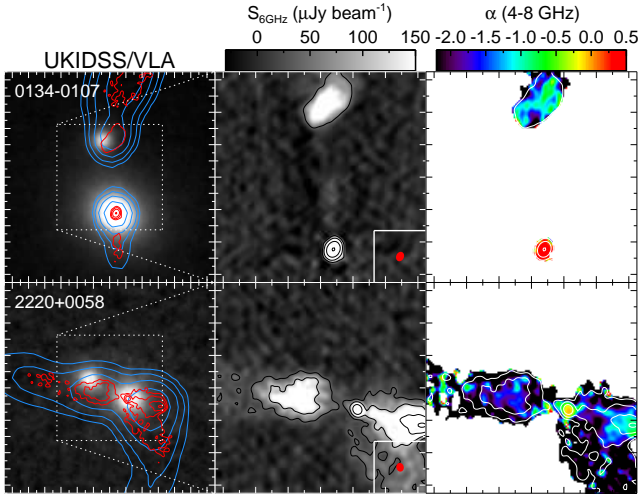}
\caption{Single FR\,II AGNs that were mistaken for binaries in the lower resolution VLA-Stripe82 maps. For each system, we show: ({\it left}) the UKIDSS $J$-band images overlaid with our VLA 6\,GHz A-configuration continuum map ({\it red} contours) and the 1.4\,GHz continuum map from the VLA-Stripe82 survey \citep[{\it blue} contours;][]{Hodge11}, ({\it middle}) a zoomed-in version of the 6\,GHz continuum map and its restoring beam (the red ellipse at the lower right corner), and ({\it right}) the 4$-$8~GHz spectral index maps. The lowest blue and red contours are at 2$\sigma$ and 4$\sigma$, respectively, and the levels increase exponentially to the peak value of each map. Major tickmarks are spaced in 1\arcsec\ intervals in all panels.
\label{fig:singles}}
\end{figure}

\begin{figure*}[!t]
\plotone{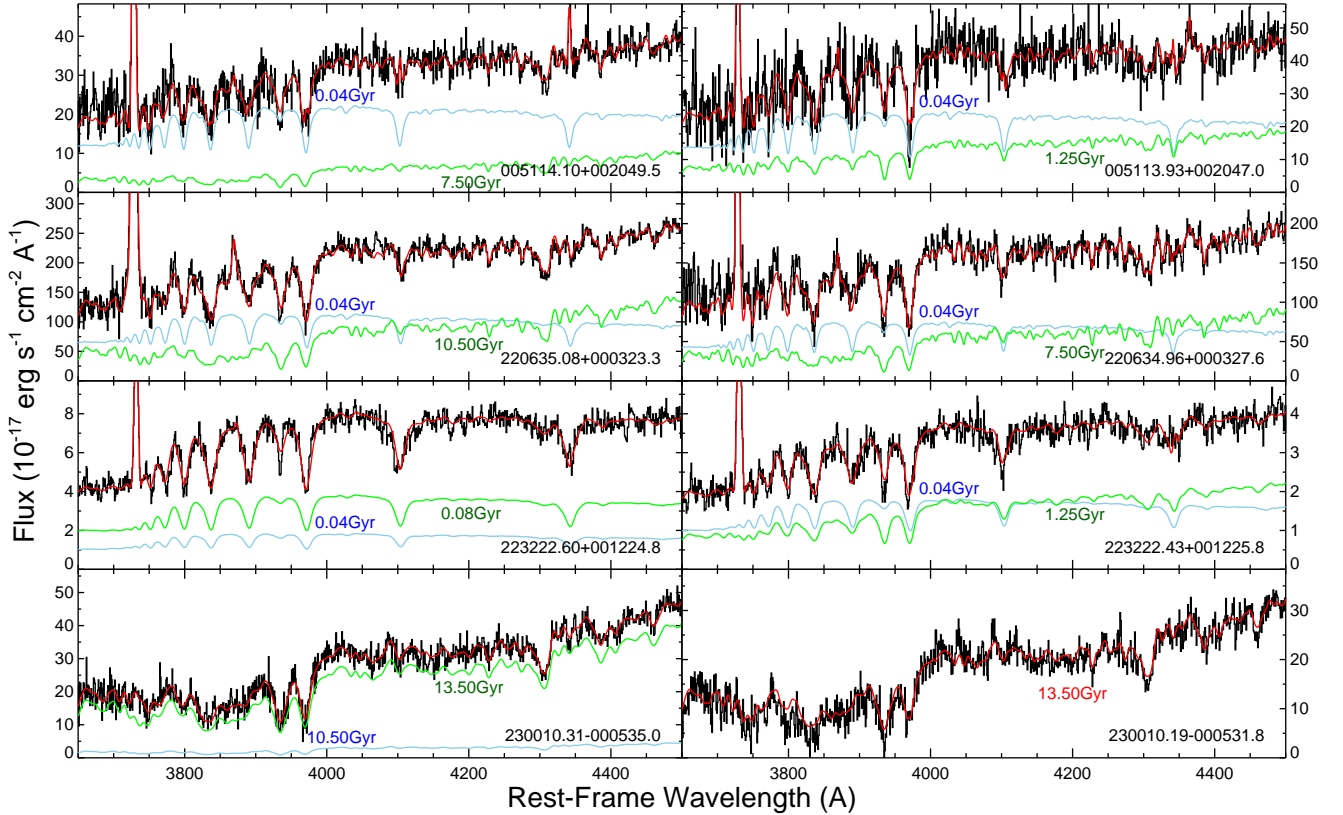}
\caption{Optical spectra of the binary AGN hosts. For clarity, we only show the spectral range between rest-frame 3650 and 4500\,\AA, which is sensitive to young stellar populations. The red curves show our best-fit SSP$+$emission-line models. The blue and green curves show the two SSPs that dominate the luminosity in this wavelength range; their ages are labeled. A single 13.5~Gyr SSP fits the spectrum of 230010.19$-$000531.8.  
\label{fig:specs}}
\end{figure*}

\begin{deluxetable}{lccccc}
\small
\tablewidth{0pt}
\tablecaption{VLA A-configuration C-band Observations
\label{tab:obs}}
\tablehead{ 
\colhead{Target} & \colhead{UT Date} & \colhead{Int Time}  & \colhead{rms} & \colhead{Beam} & \colhead{PA} \\
\colhead{} & \colhead{} & \colhead{min} & \colhead{\uJy/bm} & \colhead{arcsec} & \colhead{deg}
}
\startdata
\multicolumn{6}{c}{Confirmed Candidates} \nl
0051$+$0020 & 2015 Jul 06 & 60.0 & 4.6 & 0.27$\times$0.24 & $-$13.8 \\
2206$+$0003 & 2015 Jun 30 & 58.0 & 4.4 & 0.27$\times$0.23 & $+$19.9 \\
2232$+$0012 & 2015 Jun 30 & 58.1 & 4.2 & 0.27$\times$0.23 & $+$16.1 \\
2300$-$0005 & 2015 Jul 05 & 57.6 & 5.1 & 0.31$\times$0.23 & $+$32.1\\
\multicolumn{6}{c}{Rejected Candidates} \nl
0134$-$0107 & 2015 Jul 06 & 20.8 &10.5 & 0.30$\times$0.24 & $-$24.0 \\
2220$+$0058 & 2015 Jun 30 & 16.4 & 8.5 & 0.27$\times$0.23 & $+$15.0 
\enddata
\end{deluxetable}

We observed six spectroscopically confirmed candidates from \citet{Fu15} with the VLA in the A configuration with the C-band receivers\footnote{program 15A-211 (PI: Fu)}. The sample includes five grade-A and one grade-B candidates (2220$+$0058). The receivers have a total bandwidth of 4\,GHz at a central frequency of 5.9985\,GHz. We grouped the targets into three scheduling blocks (SBs) of 1.8 to 2.8\,hours. A nearby unresolved calibrator was observed every $\sim$10~min. At the beginning of each SB, 3C\,48 was observed for bandpass and flux-density calibration. Table\,\ref{tab:obs} summarizes the observations.

The observations were calibrated using the Common Astronomical Software Applications (CASA) package. We used the VLA pipeline to perform basic flagging and calibration. Additional flagging was deemed unnecessary after inspecting the data quality. We used the self-calibration technique to improve the calibration derived from the calibrator for 0134$-$0107. For imaging and deconvolution with the CASA task {\sc clean}, we used a Briggs robust parameter of $0$ to achieve good image fidelity with only a modest loss of sensitivity compared to ``natural'' weighting. The resulting restoring beams are on average 0\farcs28$\times$0\farcs23 in FWHM (Table\,\ref{tab:obs}). We used a pixel size of 0\farcs05 for the images to sample the FWHM of the restoring beam with $\sim$5 pixels. We ran the iterative clean process until the residuals dropped below 3$\times$ the expected radiometer noise (4.2--12\,\uJy\,beam$^{-1}$). To model the frequency dependence of the emission within the 4\,GHz bandwidth, we adopted two Taylor terms with {\sc clean}. We then computed the spectral index map using the ratio of the two Taylor terms for regions with sufficient S/N \citep{Rau11}. 

\section{Analysis and Results} \label{sec:analysis}

\begin{deluxetable*}{lcccccccccc}
\tablewidth{0pt}
\tablecaption{Optical and Radio Properties of the Confirmed Binaries
\label{tab:data}}
\tablehead{ 
\colhead{Radio Designation} & \colhead{Mode} & \colhead{$z$} & \colhead{$M_{\star}$} & \colhead{$\sigma_{\star}$} & \colhead{log($\lambda$)} & \colhead{$S^{\rm tot}_{\rm 6 GHz}$} & \colhead{$\alpha_{\rm 4-8GHz}$} & \colhead{Maj} & \colhead{Min} & \colhead{P.A.} \\
\colhead{J2000} & \colhead{} & \colhead{} & \colhead{log(\msun)} & \colhead{\kms} &  \colhead{} & \colhead{mJy} & \colhead{} & \colhead{arcsec} & \colhead{arcsec} & \colhead{deg} \\
\colhead{(1)} & \colhead{(2)} & \colhead{(3)} & \colhead{(4)} & \colhead{(5)} & \colhead{(6)} & \colhead{(7)} &  \colhead{(8)} & \colhead{(9)} & \colhead{(10)} & \colhead{(11)}  
}
\startdata
005113.93$+$002047.2&HE&0.11253& 10.8&  140$\pm$13  & $-$1.0$\pm$0.4 &0.296$\pm$0.012&$-$1.08$\pm$0.22&0.083$\pm$0.020&0.060$\pm$0.026&$+$22$\pm$31\\
005114.11$+$002049.5&HE&0.11257& 10.9&  166$\pm$9   & $-$1.3$\pm$0.4 &0.934$\pm$0.014&$-$0.93$\pm$0.09&0.208$\pm$0.005&0.196$\pm$0.005&$-$31$\pm$15\\
220634.98$+$000327.6&HE&0.04656& 10.3&  115$\pm$9   & $-$1.4$\pm$0.4 &0.081$\pm$0.020&$-$1.69$\pm$1.67&0.414$\pm$0.105&0.192$\pm$0.076&$+$73$\pm$14\\
220635.08$+$000323.2&HE&0.04640& 10.7&  171$\pm$6   & $-$1.5$\pm$0.4 &0.581$\pm$0.013&$-$1.14$\pm$0.12&0.161$\pm$0.006&0.104$\pm$0.011&$-$75$\pm$6 \\
223222.44$+$001225.9&HE&0.22128& 10.6&  245$\pm$37  & $-$2.9$\pm$0.4 &0.139$\pm$0.014&$-$1.80$\pm$0.73&0.387$\pm$0.038&       $<$0.069&$-$71$\pm$4 \\
223222.60$+$001224.7&HE&0.22187& 10.7&  211$\pm$61  & $-$2.0$\pm$0.4 &0.243$\pm$0.010&$-$1.29$\pm$0.30&0.148$\pm$0.018&       $<$0.087&$+$47$\pm$10\\
230010.18$-$000531.6&LE&0.17971& 11.4&  285$\pm$13  & $-$4.0$\pm$0.7 &0.512$\pm$0.012&$-$0.52$\pm$0.14&       $<$0.174&       $<$0.036&  \nodata \\
230010.24$-$000533.9&LE&0.17981& 11.5&  324$\pm$12  & $-$4.4$\pm$0.7 &0.068$\pm$0.014&$-$1.16$\pm$1.44&       $<$0.663&       $<$0.111&  
\nodata
\enddata
\tablecomments{
Rows are grouped in sets of two that each include a pair, and sources are sorted in ascending RA.
(1) J2000 coordinates of the centroid of the radio source from our VLA C-band data;
(2) mode of AGN activity (HERG or LERG, \S\ref{sec:RAGNs}); 
(3) spectroscopic redshift from stellar absorption features;
(4) total stellar mass measured from modeling the optical spectrum with simple stellar populations (\S\ref{sec:host});
(5) intrinsic velocity dispersion of the stellar populations;
(6) Eddington ratio, $\lambda = (L_{\rm rad} + L_{\rm mech})/L_{\rm Edd}$ (\S\ref{sec:BHAR}); 
(7) total flux density at 6~GHz from our VLA C-band data;
(8) spectral index within the 4~GHz bandwidth of the C-band data;
(9-11) beam-deconvolved source sizes (FWHMs in arcsec) along the major and minor axes and the position angle of the major axis (degrees east of north); 3$\sigma$ upper limits are given for unresolved sources. The 1$\sigma$ error bars in Columns 7$-$11 are derived from Monte Carlo simulations (\S\ref{sec:RAGNs}). 
The error of the 6~GHz photometry does not include the 3\% uncertainty in the VLA flux density scale \citep{Perley13}.
}
\end{deluxetable*}

By comparing our radio maps to SDSS optical images and UKIRT Infrared Deep Sky Survey (UKIDSS) near-IR images, we confirm that four of the five grade-A candidates are binary AGNs (Fig.\ \ref{fig:binaries}). All are major mergers with stellar mass ratios less than 3:1 (primary/secondary). As we expected from the lower resolution VLA-Stripe82 data, each nucleus in the confirmed binaries is associated with a compact steep-spectrum radio source. The radio positions are consistent with the centroids from SDSS and UKIDSS images; i.e., we do not detect any significant spatial offset between the AGNs and the stellar nuclei. These properties are similar to the VLA-confirmed binary AGN SDSS~J1502$+$1115 at $z = 0.39$ \citep{Fu11b}. \citet{Muller-Sanchez15} identified three double-radio-core sources from a VLA survey of 18 double-peaked [O\,{\sc iii}] AGNs. The radio sources show flat spectrum ($\alpha \lesssim -0.5$) and are associated with spatially resolved emission-line components within a single stellar component. The flat radio spectrum could be due to either synchrotron self-absorption in dense AGN-energized cores or free-free absorption in the ionized diffuse gas. If the former, these binary AGNs may represent a later merger stage than our sources because their stellar bulges may have coalesced. 

The remaining grade-A candidate (0134$-$0107) and the only grade-B candidate (2220$+$0058) are both single Fanaroff-Riley\,II (FR\,II) radio galaxies involved in galaxy mergers (Fig.~\ref{fig:singles}). They were mis-classified as binaries because their radio jets/lobes align with their merging companions. These contaminating sources are difficult to remove without sub-arcsec resolution radio maps. If these are not due to projection effects, it is possible that their radio jets interacted with gas in the companion galaxies and formed the hot spots near the companion galaxies. These FR\,II sources would be interesting laboratories to study jet-cloud interactions, and they may lend insight into whether radio outflows from an AGN could suppress star formation in not only the AGN's host galaxy but also in companion galaxies.

\subsection{Radio Emission} \label{sec:RAGNs}

The radio sources in the four confirmed binaries have been classified as AGNs in \citet{Fu15} because of either (1) their excess of radio power vs.\ extinction-corrected H$\alpha$ luminosity (2232$+$0012 and 2300$-$0005\footnote{Although the galaxies in 0051$+$0020 and 2206$+$0003 are not formally classified as radio-excess objects, their radio powers are still several times greater than that implied by their H$\alpha$-traced SFRs \citep{Fu15}.}), and/or (2) their location on the \citet[][]{Baldwin81} [N\,{\sc ii}]/H$\alpha$ diagram (0051$+$0020, 2206$+$0003, and 2232$+$0012\footnote{0051$+$0020 and 2206$+$0003 were first identified as candidate binary AGNs from their SDSS spectra by \citet{Liu11}; 220635.08+000323.2 is classified as a LINER and the other five galaxies are classified as AGN-starforming composite galaxies \citep{Fu15}. Galaxies in 2300$-$0005 do not show detectable emission lines; therefore, they cannot be classified based on the BPT diagram.}). We further classify the AGNs into high-excitation radio galaxies (HERGs; radiative/quasar/cold-mode AGNs) and low-excitation radio galaxies (LERGs; jet/radio/hot-mode AGNs) \citep{Hine79,Laing94} based on their emission-line ratios and strengths using the composite, multi-step classification scheme of \citet{Best12} (Table~\ref{tab:data}). Interestingly, the accretion modes are the same for both components in all four binaries. The four binary AGNs have 1.4\,GHz radio powers between $1.0\times10^{22}$ and $1.7\times10^{23}$\,W\,Hz$^{-1}$ \citep{Fu15}. Similarly low radio-luminosity AGNs are dominated by LERGs \citep[e.g.,][]{Best12}, yet six out of eight AGNs in the binaries are HERGs with strong optical emission lines and only two are LERGs with weak emission lines. This suggests interactions may preferentially trigger cold-mode accretion, but a larger sample is clearly needed to draw any solid conclusion on this matter. 

The AGN nature is also supported by the compact radio morphologies (Fig.~\ref{fig:binaries} and Table~\ref{tab:data}). To measure the radio source parameters, such as size, flux density, and spectral index, we fit an elliptical Gaussian to each component in the 6\,GHz Stokes {\it I} map. We then deconvolve the restoring beam to estimate the intrinsic size and orientation of the radio sources. Finally, we use the best-fit model and the spectral index map to measure the flux-weighted spectral indices for each component. Because pixels are correlated in interferometer images, we use Monte Carlo simulations to estimate the uncertainties of each parameter. Once a best-fit model is obtained for a given source, we generate 300 synthetic maps with the same source and noise properties. We measure the synthetic source parameters with elliptical Gaussians as for the real data. Dual-channel synthetic datacubes centered at 5 and 7\,GHz are generated in order to measure the spectral indices. The dispersion of the parameter measurements across the synthetic sample are quoted as the 1$\sigma$ errors. The cross-band 1.4\,GHz-to-6\,GHz spectral indices are less reliable because of the mismatch in spatial resolutions, but they are consistent with the intra-band spectral slopes.  

Table~\ref{tab:data} lists the best-fit parameters and their associated uncertainties for the confirmed binaries. All eight radio sources are compact ($\lesssim$0\farcs4 or $\lesssim$0.7\,kpc at $z = 0.1$) and show steep spectra ($\alpha_{\rm 4-8GHz} \lesssim -0.5$). These properties are consistent with synchrotron emission from small-scale radio jets and are similar to SDSS~J1502$+$1115 \citep{Fu11b}.

\subsection{Host Galaxies} \label{sec:host}

We model the optical spectra to measure the properties of the host galaxies and the AGN narrow-line regions. We use the SDSS spectra for 0051+0020, 2206+0003 and 2300-0005, and the Keck/LRIS spectra for 2232+0012 \citep{Fu15}. We model each spectrum as a number of emission lines and a weighted sum of simple stellar populations (SSPs), convolved by the line-of-sight velocity distribution (LOSVD) and reddened in rest-frame by a Calzetti extinction law \citep{Calzetti00}. The LOSVD and the profile of the emission lines are parameterized as two independent Gauss-Hermite series \citep{van-der-Marel93} to the fourth order. Both the SSP templates and the Gaussian emission-line templates are convolved to match the wavelength-dependent spectral resolution of each spectrum. We used the Solar-metallicity SSP templates of MIUSCAT \citep{Vazdekis12}, which have a wide spectral range (3465--9469\,\AA) with a uniform spectral resolution (FWHM = 2.5\,\AA). 

Table~\ref{tab:data} lists the redshifts from stellar absorption features, the total stellar mass (corrected for extinction and aperture loss), and the stellar velocity dispersions. These are all massive galaxies with stellar masses between $10.3 < {\rm log}(M_\star/M_\odot) < 11.5$, although the two LERGs are distinctly more massive than the HERGs, consistent with the general radio-AGN population \citep{Best12}. The stellar velocity dispersions are in the range $120 < \sigma_\star < 320$\,\kms, implying black hole masses in the range $7.4 < {\rm log}(M_\bullet/M_\odot) < 9.4$ \citep{Kormendy13}. The mass-weighted stellar ages are old ($>$\,9\,Gyr) for all except for the two galaxies in 2232$+$0012 ($\sim$\,3\,Gyr). However, the presence of Balmer absorption lines in the spectra of the six HERGs shows a significant contribution from a young ($\sim$\,40\,Myr) stellar population (Fig.\ \ref{fig:specs}). This agrees with the AGN-starforming composite line ratios, suggesting active star formation in the host galaxies. The spectra of the two LERGs in 2300$-$0005 are consistent with purely old ($\gtrsim$\,10\,Gyr) stellar populations. The merger in 2300$-$0005 apparently activated black hole accretion without triggering any visible recent star formation. 

\subsection{Black Hole Accretion Rate} \label{sec:BHAR}

To estimate the Eddington-scaled accretion rate we first obtain the bolometric radiative luminosity using the reddening-corrected \OIII\ luminosity \citep[$L_{\rm rad} = 3500 L_{\rm OIII}$;][]{Heckman04}. We then derive the jet mechanical luminosity from the 1.4\,GHz radio luminosity \citep[$L_{\rm mech} = 7.3\times10^{43}{\rm\,erg\,s}^{-1} (L_{\rm 1.4GHz}/10^{24}{\rm\,W\,Hz}^{-1})^{0.7}$;][]{Cavagnolo10}. Finally, we estimate the black hole mass from the stellar velocity dispersion \citep[$M_\bullet = 3.1\times10^8\,M_\odot (\sigma_\star/200{\rm\,km\,s}^{-1})^{4.38}$;][]{Kormendy13} but caution that the uncertainty of $M_\bullet$ is $\sim$0.7\,dex because of the $\sigma_\star$ oscillation observed in merger simulations \citep{Stickley12}. The Eddington luminosity is defined as $L_{\rm Edd} = 1.26\times10^{38} M_\bullet/M_\odot$~erg~s$^{-1}$ for pure hydrogen, and the Eddington ratio is $\lambda = (L_{\rm rad} + L_{\rm mech})/L_{\rm Edd}$. The uncertainty of the $\lambda$ estimates is dominated by the uncertainties of the empirical calibrations for $L_{\rm rad}$ and $L_{\rm mech}$ and is $\sim$0.4~dex for HERGs and $\sim$0.7~dex for LERGs. The binary AGNs show similar Eddington ratios to the general radio-AGN population at low redshift \citep[e.g.,][]{Best12}: the six HERGs have significantly higher Eddington ratios ($-2.9 <$ log($\lambda$) $< -1.0$) than the two LERGs (log($\lambda$) $\lesssim -4.0$; Table~\ref{tab:data}). 

\subsection{Duty Cycle of Synchronized Accretion}

We estimate the duty cycle of synchronized accretion using the binary fraction of radio-AGNs in major mergers. Keep in mind that we have included both low- and high-excitation modes and the majority of the binary AGNs are HERGs. We focus on the 715 radio sources in \citet{Hodge11} that have $0.005<z_{\rm spec}<0.2$ from SDSS DR10 so that both the binary sample and the control sample have the same spectroscopic incompleteness. Objects with Keck spectra are excluded in this discussion because of their different sample bias. The redshift range matches that of the three confirmed binary AGNs with SDSS spectra (0051$+$0020, 2206$+$0003, 2300$-$0005)\footnote{The other binary, 2232$+$0012 at $z = 0.22$, only has Keck spectra.}. We find that 551 of the 715 sources are classified as AGNs because of the radio excess and/or the emission-line ratios. But only 13 (2.4\%) of these AGNs have companion galaxies with angular separations less than 5\arcsec, projected separations less than 10\,kpc, and comparable $r$-band magnitudes ($\Delta r < 1.2$ mag). These are either major mergers or projected galaxy pairs, and they include the three confirmed binary AGNs and one grade-B binary candidate (0149$-$0014; \citealt{Fu15}). In comparison, we find 215 such pairs in the 92~deg$^2$ VLA-Stripe82 area using the 22,192 SDSS spectroscopic targets between $0.005<z_{\rm spec}<0.2$. Spurious pairs were removed through visual inspection. This result implies that (1) radio-AGN occurs $\sim$6\% (13/215) of the time in a major merger when the physical separation is below 10~kpc, and (2) synchronized black hole accretion occurs $\geqslant$$23^{+15}_{-8}$\% (3/13) of the time when such a merger is detectable as a radio-AGN because of one of the galaxies. If black hole accretion happens stochastically instead of being triggered by merger, one would expect $\sim$23$\times$ fewer binary AGNs, because the average AGN duty cycle is $\lesssim$1\% at $z<0.5$ \citep[e.g.,][]{Shankar09}. The observed high duty cycle of synchronized accretion thus strongly suggests an AGN-merger connection, and it is in general agreement with smoothed particle hydrodynamics merger simulations adopting the Bondi-Hoyle accretion formula \citep{Van-Wassenhove12,Blecha13a}.

We can also estimate the average AGN lifetime and the duration of the binary phase. A major merger can be observed as a galaxy pair over the dynamical friction timescale, which is $\sim$100~Myr using the Chandrasekhar formula \citep{Foreman09} with typical host galaxy parameters ($r_{\rm max} = 10$~kpc, $\sigma = 200$~\kms, and $M = 10^{10}$~\msun). Combined with the $\sim$6\% duty cycle for the radio-AGN phase and the $\gtrsim$23\% duty cycle for the binary AGN phase, we estimate that the average lifetime for a single AGN is $\sim$3.7~Myr ($=100{\rm\,Myr}\times0.06\times(1+0.23)/2$) and the synchronized accretion phase could last $\gtrsim$1.4~Myr ($=100{\rm\,Myr}\times0.06\times0.23$). Of course, all of the above are subject to the luminosity limit of the VLA-Stripe82 survey ($L_{\rm 1.4GHz} \gtrsim 10^{22}$ W~Hz$^{-1}$ at $z = 0.1$).

\section{Summary} \label{sec:discussion}

We have conducted high-resolution deep VLA continuum observations of six kpc-scale major galaxy mergers that are associated with double radio sources. These sources are strong candidate binary AGNs, for which the simultaneous accretion could be triggered by tidal interactions. The new 0\farcs3 observations confirmed four binary AGNs and rejected two others. This confirmation rate is quite high, especially when only considering the five grade-A candidates, four of which are binaries. We studied the properties of the AGNs and their host galaxies in the four confirmed binaries. We found that all of the AGNs are compact, steep-spectrum sources at 6\,GHz. Six are HERGs that are hosted by $2-8\times10^{10}$~\msun\ galaxies and that are accreting at 0.1 to 10\% of the Eddington limit, and two are LERGs that are hosted by $\sim3\times10^{11}$\,\msun\ galaxies and that are accreting at $\lesssim$0.01\% of the Eddington limit. A significant young stellar population is evident in the optical spectra of the HERGs, but only old stellar populations are found in the LERGs. These properties are in agreement with those of the general radio-AGN population at low redshifts. The confirmation of this population of kpc-scale binary AGNs indicates a $\geqslant$23\% duty cycle of synchronized black hole accretion in major mergers. Our result is consistent with simulations that suggest an offset in AGN triggering times in asymmetric mergers and indicates that mergers trigger and tend to synchronize AGN activity at small separations. 

\acknowledgments
We thank Cornelia Liang, Robert Mutel, Steve Spangler, and Alan Stockton for helpful discussions and the referee for useful comments.
HF was partially supported by NASA JPL award 1495624 and University of Iowa funds. 
ADM was partially supported by NASA ADAP award NNX12AE38G and by NSF awards 1211112 and 1515404.
SGD acknowledges support from NSF award AST-1413600.
The National Radio Astronomy Observatory is a facility of the National Science Foundation operated under cooperative agreement by Associated Universities, Inc.

{\it Facilities}: VLA, Sloan, UKIDSS, Keck:I


\end{document}